\documentclass[aip,rsi,graphicx]{revtex4-1}
\usepackage[latin1]{inputenc}
\usepackage{amsmath}
\usepackage{amsfonts}
\usepackage{amssymb}
\usepackage{graphicx}
\usepackage{fullpage}
\usepackage{float}
\usepackage[parfill]{parskip} 
\usepackage{color}
\usepackage[colorlinks=true, pdfstartview=FitV, linkcolor=blue,citecolor=blue, urlcolor=blue]{hyperref}
\usepackage{dcolumn}%
\usepackage{bm}%

\draft 

\begin{document}


\title{Ultrahigh Transmission Optical Nanofibers} 



\author{J. E. Hoffman,$^1$ S. Ravets,$^{1, 2}$ J. A. Grover,$^1$ P. Solano,$^1$  P.  R. Kordell,$^1$ J. D. Wong-Campos,$^1$  L. A. Orozco,$^1$ S. L. Rolston}
\affiliation{$^1$Joint Quantum Institute, Department of Physics, University of Maryland, and National Institute of Standards and Technology, College Park, MD 20742, U. S. A.\\
$^2$ Laboratoire Charles Fabry, Institut d'Optique, CNRS Univ Paris-Sud, Campus Polytechnique, RD 128, 91127 Palaiseau cedex, France  \\
$^*$Corresponding author: rolston@umd.edu}


\date{\today}

\begin{abstract}
We present a procedure for reproducibly fabricating ultrahigh transmission optical nanofibers (530 nm diameter and 84 mm stretch) with single-mode transmissions of 99.95 $ \pm$ 0.02 \%, which represents a loss from tapering of 2.6 $\,\times \,$ 10$^{-5}$ dB/mm when normalized to the entire stretch.  When controllably launching the next family of higher-order modes on a fiber with 195 mm stretch, we achieve a transmission of 97.8 $\pm$ 2.8\%, which has a loss from tapering of 5.0 $\,\times \,$ 10$^{-4}$ dB/mm when normalized to the entire stretch.  Our pulling and transfer procedures allow us to fabricate optical nanofibers that transmit more than 400 mW in high vacuum conditions.  These results, published as parameters in our previous work, present an improvement of two orders of magnitude less loss for the fundamental mode and an increase in transmission of more than 300\% for higher-order modes, when following the protocols detailed in this paper.   We extract from the transmission during the pull, the only reported spectrogram of a fundamental mode launch that does not include excitation to asymmetric modes; in stark contrast to a pull in which our cleaning protocol is not followed.  These results depend critically on the pre-pull cleanliness and when properly following our pulling protocols are in excellent agreement with simulations.
\end{abstract}


\maketitle 


\section{Introduction}
Optical nanofibers have seen widespread use in science and engineering applications in the last thirty years\cite{Brambilla2010, Morrissey2013}.  The tight confinement of the evanescent field around the optical nanofiber \cite{Kien2004b}, unique light geometries provided by the fiber modes \cite{Sague2008, LeKien2004, Reitz2012}, low loss, and promise of improved atom-light interaction \cite{Alton2011, LeKien2005, Vetsch2010, Goban2012, Wuttke2012}  have led to increased interest in the physics community. Optical micro- or nanofibers are used for sensing and detection \cite{Nayak2007, Knight1997}, and coupling light to resonators \cite{Knight1997, Kakarantzas2002, Spillane2003, Louyer2005, Morrissey2009, Fujiwara2012}, NV centers\cite{Schroder2012}, or  photonic crystals\cite{Nayak2013,Thompson2013,Sadgrove2013}. Optical nanofiber fabricated systems can be connected to an existing fiber network to provide applications in quantum information science\cite{Kimble2008}. The development of atom traps around optical nanofibers affords new avenues of research \cite{LeKien2004, Vetsch2010, Goban2012} including loss-limited hybrid quantum systems. These applications can benefit from high transmission nanofibers through a reduction in unwanted stray light fields produced from non adiabatic mode excitation and reduced laser power requirements.  Here we present the tools and procedures necessary to create ultrahigh transmission nanofibers.

Previous work has focused on the post-pull environmental controls  in humidity and  air purity\cite{Fujiwara2011}; here we focus on the critical pre-pull steps necessary to achieve an ultrahigh transmission before handling the known environmental effects.  Following the protocols and procedures we have produced fibers with 99.95 \% transmission when launching the fundamental mode.  We have also launched higher-order modes\cite{Ravets2013, Ravets2013a}, where we achieve transmissions of greater than 97\% for the first family of excited modes.  This level of transmission has required a thorough optimization of the pulling algorithm and of the cleaning procedure. Here, we explicitly explain our technique and our apparatus that produces fibers that we will use in a series of experiments in quantum hybrid systems \cite{Hoffman2011, Hafezi2012}.

Our pulling technique, using a well-known methodology\cite{Brambilla2010}, requires two pulling motors and a stationary oxyhydrogen heat source. The flame brushing method allows us to reliably produce optical nanofibers with controllable taper geometries and a uniform waist.  The waist can vary in length from 1 to 100 mm, and we can achieve radii as small as hundreds of nanometers \cite{Bilodeau88, Birks1992, Warken2007,Warken2007b, Garcia-Fernandez2011}.   Rather than sweep the flame back and forth over the fiber, we keep the flame stationary; this action prevents the creation of small air currents, which could lead to nonuniformities on the fiber waist and is equivalent to transforming to the rest frame of the flame.  This transformation is applicable to other pulling techniques as well, so there would be no need to scan a heat source.   

Other common techniques for optical nanofiber production include micro-furnaces, fusion splicers,  chemical etchants, and CO$_2$ lasers \cite{Ding2010, Lambelet1998, Kbashi2012, Yokota1997, Dimmick1999, Ward2006}.  Chemical etching generally produces lower transmission than other heat and pull methods and offers less control over the shape of the taper and the length of the waist.   A CO$_2$ laser produces high-transmission optical nanofibers but the final diameter is limited by the power and focus of the laser.  Here we present, to best of our knowledge, the highest recorded transmission optical nanofibers, with a loss of 2.6 $\times\,10^{-5}$~dB/mm on the fundamental mode \cite{Brambilla2010, Stiebeiner2010}, with controllable taper geometries and long fiber waists which are suitable for the space and optical constraints of cryogenic environments \cite{Hoffman2011}. 

The paper is organized as follows:  Section~\ref{sec:exp} discusses the experimental setup.   In Sec.~\ref{sec:pp} we describe the necessary steps for a typical pull, detailing the cleaning and alignment process. Sec.~\ref{sec:res} examines the quality of the produced fibers by measuring their transmission and comparing our results to other reported measurements.  In Sec.~\ref{sec:PM} we report the transmission of more than 400 mW through the nanofiber under high vacuum conditions.  Sec.~\ref{sec:con} presents our conclusions.   Appendices \ref{sec:alg}  and \ref{sec:dustaccumulation} detail procedures and controls.

\section{Experimental Setup}\label{sec:exp}

This sections describes the details of the experimental setup.  Our work follows the originally Mainz and currently Vienna group\cite{Warken2007b}.  A more detailed discussion of the algorithm and experimental verification can be found in \ref{sec:alg}.

\subsection{The fiber-pulling apparatus}
\begin{figure}[H]
\begin{center}
\includegraphics[width=\linewidth]{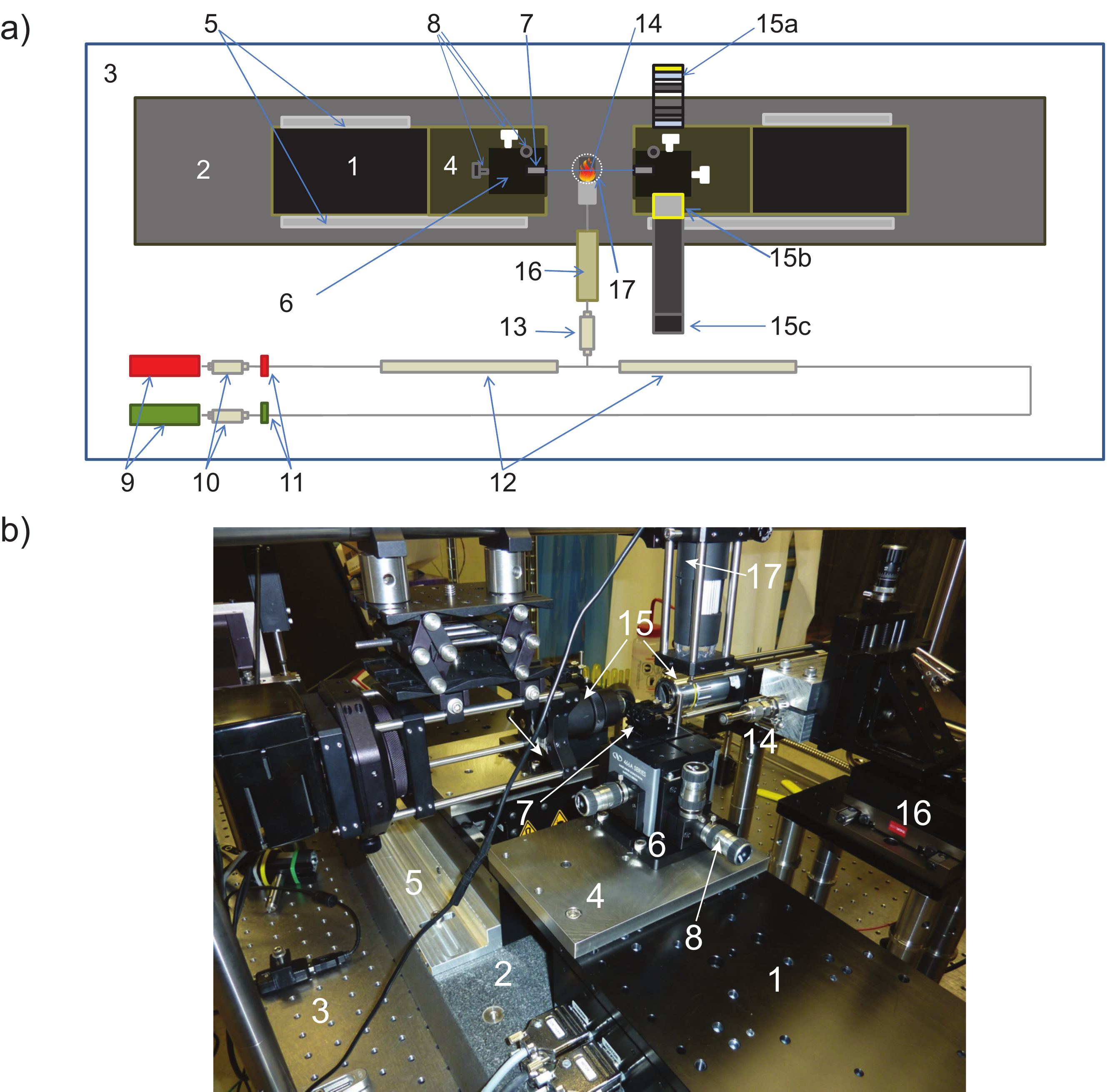}
\caption{a) A schematic of the fiber pulling setup (view from above).  b)  A photograph of the pulling apparatus with prominent pieces numbered as in Table~\ref{tab:Table}.
1) Fiber motors.
2)  Granite slab.
3)  Optical breadboard.
4)  Adapter plates.
5)  L brackets.
6)  XYZ fiber alignment flexure stages.
7) Fiber holders.
8)  Adjustment screws.
9) Flow meters.
10)  Filters. 
11) Valves.
12) Pipes.
13)  Filter.
14) Nozzle. 
15) (a) Illumination system, (b) Optical microscope, and  (c) CCD.
16) Flame positioning stepper motor. 
17) 2 MP USB microscope positioned orthogonally to the fiber.
The entire apparatus is inside a cleanroom rated to ISO Class 100.}
\label{fig:pullingrig}
\end{center}
\end{figure}

The fiber-puller apparatus (see. Fig.~\ref{fig:pullingrig} and Table~\ref{tab:Table}) consists  of a heat source that brings the glass to a temperature greater than its softening point (1585$^{\circ}$ C for fused silica \cite{Corning}) and two motors that pull the fiber from both ends. We use two computer-controlled motors, Newport XML 210 (fiber motors),  mounted to a precision-ground granite slab with dimensions  12"~$\times$~48"~$\times$~4", flat to 3.81 $\mu$m on average.  The granite slab serves two purposes: it is a damping weight and a flat surface. The weight of the granite slab, exceeding 100 kg, damps the recoil from the fiber motors as they change direction at the end of every pull step.  Without a flat surface the motors will not work to specification, leading to parasitic effects on the nanofiber: the pitch or yaw of the motor can vary the distance between the fiber and the flame, changing the effective size of the flame and pulling the fiber in various directions (negating any pre-pull alignment).  The motors  are mounted to the granite by L-bracket adapters designed to not deform the motors from the the granite surface.  We then mount the granite to an optical breadboard on three points so that surface imperfections of the optical table do not distort the granite slab.

The fiber motors, have 210 mm of motion, with a minimum incremental motion of 0.01 $\mu$m and an on-axis accuracy of 3 $\pm \ 1.5\ \mu$m (1 in Fig.~\ref{fig:pullingrig}).  The resolution of the motors is much smaller than any other length scale in our system, making the motors suitable for pulling.  The fiber motors are controlled with a Newport XPS controller, which allows us to implement trajectories with a constant acceleration, resulting in jerkless motion during each step of the pull.
\begin{table}[H]
\begin{center}
\begin{tabular}{l l l}
\hline \hline
Item    & Part   & Description \\ \hline 
1 & Newport XML 210 & Computer-controlled high precision motor \\ 
2 & Granite slab& 12" $\times$ 48" $\times$ 4", Flat to 3.81 $\mu m$ \\
3 & Newport VH3660W & 3' x 5' workstation \\  
4 &Adapter Plate & Adapts metric XML 210 to 466A \\ 
5 & L brackets & Adapts XML 210 to granite\\ 
6 & Newport 466A &  Compact XYZ fiber alignment flexure stages \\ 
7&Newport 466A-710 & Double arm bare fiber holder double V-grooves\\ 
8 & Newport DS-4F& High precision adjuster, 8.0 mm coarse, 0.3 mm fine travel\\ 
8 &Newport AJS100-0.5   & High precision small knob adjustment screw, 12.7 mm travel\\ 
9 &Omega FMA 5400/5500 & Flow meters \\ 
10 & Swagelok SS-4F-7& Particulate filter,  7 micron pore size \\ 
11& Swagelok SS-4P4T & Valve to close the flow of gas\\ 
12 & Swagelok SS-FM4SL4SL4-12 & Stainless steel flexible tubing\\ 
13 & GLFPF3000VMM4& ``Mini Gaskleen filter" from Pall, removal rating: $\geq$ 0.003 $\mu$m\\ 
14 & Stainless Steel Custom flame nozzle &  29, 228 $\mu$m holes in a 1x2 mm$^2$ array \\ 
15 & Optical microscope & Microscope objective, CCD, and illumination system \\ 
15a & Illumintation System & Kohler illumination system \\ 
15b& Mitutoyo M Plan APO 10X&  Microscope objective, 0.28 NA, working distance 33.5 mm \\ 
15c &Flea2G  CCD camera & 2448 x 2048 pixels, 3.45 x 3.45 $\mu$m$^2$ pixels\\ 
16 & Thorlabs DRV014 & 50 mm Trapezoidal Stepper Motor Drive\\
17 & USB microscope & 200x, 2 MP USB microscope \\ 
18 & Platinum wire & Platinum catalyst to ignite flame \\ 
19 & Clean room & ISO class 100 cleanroom  \\
\hline \hline
\end{tabular}
\end{center}
\caption{List of equipment parts for the pulling apparatus with numbers corresponding to Fig.~\ref{fig:pullingrig}.}\label{tab:Table}
\end{table}

Attached to each XML 210 (1 in Fig.~\ref{fig:pullingrig}) are Newport 466A flexure stages (6 in Fig.~\ref{fig:pullingrig}), each with a Newport 466A-710 fiber clamp on top (7 in Fig.~\ref{fig:pullingrig}). We position the v-grooves of the fiber clamp on each stage at the minimum separation allowed by the parameters of a given pull, which is typically 3 cm.  Separating the fiber clamps at the minimum distance minimizes the fiber sag during the pull.  The v-grooves of the fiber clamps on the left and right fiber motors must be aligned within micrometers to achieve a high transmission.  We align the v-grooves  using  Newport DS-4F (8 in Fig.~\ref{fig:pullingrig}) and AJS100-0.5 (8 in Fig.~\ref{fig:pullingrig}) micrometers, attached to the flexure stages to allow for three axis translation. To perform the alignment we use an $in\ situ$  optical microscope (15(a)-(c) in Fig.~\ref{fig:pullingrig}).

The optical microscope includes  a  Mitutoyo M Plan APO 10X infinity-corrected objective and a Point Grey Flea2G CCD camera (15(a)-(c) in Fig.~\ref{fig:pullingrig}). The Flea2G has 2448$\times$2048, 3.45$\times$3.45 $\mu$m$^2$ pixels. With the inclusion of the long working distance microscope objective each pixel corresponds to 0.345$\times$0.345 $\mu$m$^2$ in the image. The microscope is illuminated by a K\"{o}hler illumination system composed of a thermal light source, two condenser lenses, and two apertures. 

We use an oxyhydrogen flame as a heat source to thin the fibers, in a stoichiometric mixture of hydrogen and oxygen to ensure that water vapor is the only byproduct. Stainless steel gas lines introduce the hydrogen (red) and oxygen (green) to two Omega FMA 5400/5500 flow meters (9 in Fig.~\ref{fig:pullingrig}).  The flow rates are set to 30 mL/min and 60 mL/min for oxygen and hydrogen respectively.  

 Directly after the flow meters is a coarse particle filter (10 in Fig.~\ref{fig:pullingrig}), followed by a  valve for safety (11 in Fig.~\ref{fig:pullingrig}).   We mix the gases in a tee after a flexible stainless steel tube (12 in Fig.~\ref{fig:pullingrig}).  The gas mixture is finely filtered with a high quality 3 nm filter (13 in Fig.~\ref{fig:pullingrig}).  Finally the hydrogen-oxygen mixture exits through a custom-made nozzle (14 in Fig.~\ref{fig:pullingrig}).  The nozzle is composed of two parts and is constructed out of stainless steel.  The first part is a 6.5 mm diameter plate with a 3.175 mm thickness.  The plate has a 1 x 2 mm$^2$ array of 29, 228 $\mu$m holes.  The long axis of the holes is perpendicular to the fiber axis.   The second part of the nozzle is an adapter in which the first plate is pressure fit into the gas line.  The outer diameter is 9.5 mm and is counter sunk 3.4 mm with an inner diameter of 6.35 mm.  This piece connects to 6.35 mm outer diameter tubing that then connects to stainless steel gas line with a  Swagelok connector. The adapter was heated, allowing it to expand, so that the plate would slide into the countersunk inner diameter.  The design diameter serves as a flame arrestor, while still allowing for the gas flow to be in the laminar regime.

We ignite the flame  using a resistively-heated platinum wire as a catalyst.  This process is clean and prevents the deposition of any particulate on the fiber from the ignition process.  The nozzle is clamped to a Thorlabs DRV014 motor (16 in Fig.~\ref{fig:pullingrig}), the flame motor, that translates the flame in front of the fiber for the duration of the pulling process.  The flame motor introduces and removes the heat source and during the pull we fix the horizontal distance between the nozzle and the front edge of the fiber at 0.5 mm, as depicted in Fig.~\ref{fig:pullingrigrealpic}.   We have found experimentally that a distance of 0.4-0.6 mm provides the proper heat distribution from our flame.  Working outside this range for our flow rates significantly reduces the reproducibility of the fabrication process.  If the flow rates were to change, the optimal working distance between the nozzle and fiber would need to be modified and it would be necessary to remeasure the effective width of the flame (see Sec.~\ref{sec:flamemeasurement}).

The entire pulling apparatus is inside a cleanroom initially specified as ISO Class 100.  If any fiber buffer remains or dust lands on the fiber at any time the transmission will degrade (see Sec.~\ref{sec:dustaccumulation}).  

\subsection{Transmission monitoring setup}
\begin{figure}[H]
\begin{center}
\includegraphics[width=\linewidth]{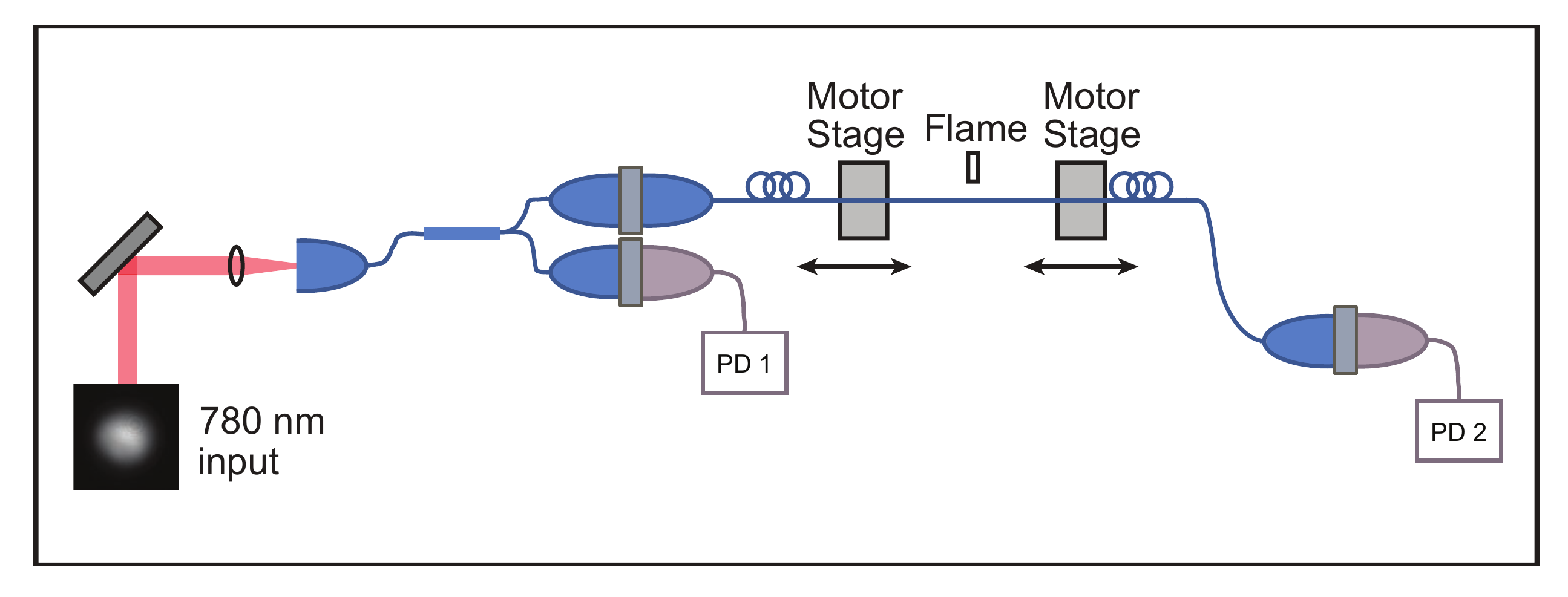}
\caption{The experimental setup to monitor the transmission when launching the fundamental mode. }
\label{fig:expsetup}
\end{center}
\end{figure}
Figure~\ref{fig:expsetup} shows the transmission monitoring setup for the fundamental mode.  Using a 780 nm Vortex laser, we launch light into a fiber and split the light with a 50/50 in-fiber beam splitter.  In one output of the beam splitter we record the laser power using a  Thorlabs DET36A photodetector.  The other output is connected to a FC connectorized fiber that we fusion splice to Fibercore SM800 fiber. We then place the SM800 fiber in the fiber puller and record the intensity of light through the fiber  at the output of the fiber puller using another DET36A for the duration of the pull. We record data on a DPO7054 Tektroniks oscilloscope in high resolution mode set to collect $10^7$ samples.  We normalize the signal through the fiber puller to the laser drift throughout the pull.

When launching higher-order modes, a superposition of the $LP_{11}$ family of modes, we do not use the in-fiber beam splitter.  Instead, we use a pick-off to track the laser drift and then free-space couple light into an SM1500 fiber from Fibercore with an initial cladding diameter of 50 $\mu$m. 

We generate the $LP_{11}$ family superposition by launching a Gaussian beam from a New Focus Vortex laser through a phase plate \cite{Fatemi2011, Pechkis2012}.  One side of the phase plate writes a $\pi$ phase shift on half of the beam.  This generates a two-lobed mode that approximates the $TEM_{01}$ free space optical mode. The $TEM_{01}$ free space optical mode is then coupled into an SM1500 fiber and excites a superposition of the $TE_{01}$, $TM_{01}$, and $HE_{21}$ modes. There is less than one percent fundamental mode corruption, which we take into consideration when calculating the transmission.

\section{The Pulling process}\label{sec:pp}

The setup for an ultrahigh transmission pull involves a series of cleaning and alignment steps.  We outline this procedure in this section.
\subsection{Cleaning procedure}\label{sec:cleaningprocedure}
Obtaining a high transmission through an optical nanofiber requires a detailed analysis of the pre-pull cleanliness of the fiber.  If any particulate remains from the fiber buffer or if dust arrives on the fiber before being introduced to the flame, the particulate will burn and greatly diminish the final transmission, see Sec.~\ref{sec:dustaccumulation}.  Furthermore, evaporate from solvents can decrease transmission.  

Our cleaning procedure starts by mechanically removing the protective plastic buffer to expose the glass of the fiber to the flame. Then we use isopropyl alcohol on lens tissue to remove larger particulate.  A few wipes of acetone\footnote{We used acetone for the data shown in this paper; however, we do not recommend its use because it can prolong the cleaning process.  SM800 fibers have a buffer made of dual acrylate, which dissolves in acetone. This is fine for chemical removal of the buffer when heated or paired with other chemicals, but when cleaning with a wipe, the acetone can spread small buffer particulate along the stripped portion of fiber, which can burn when introduced to the flame.} are then applied with class 10a cleanroom wipes from Ted Pella, in order to dissolve smaller remnants of the buffer. A final cleaning with methanol using class 10a cleanroom wipes removes any evaporate left from the previous solvents. After, we carefully lay the fiber into the grooves of the fiber clamps on the pulling apparatus and image the entire length of cleaned fiber using the optical microscope. If there is any visible dust,  particulate, or evaporate, within the 2 $\mu$m resolution of the optical microscope, we start the cleaning procedure over.  If the fiber is clean, we proceed to align it.  

\subsection{Alignment procedure}\label{alignmentprocedure}
The alignment procedure begins by properly tensioning the fiber.  We tension the fiber by moving the fiber motors apart in 200 $\mu$m increments until the fiber slides through the fiber clamps, which typically takes  800 $\mu$m of total displacement.  This allows the fiber to reach a uniform tension. 
However, early measurements showed the fiber to be overtensioned: introducing the fiber to the flame will yield immediate thinning, even if the motors are stationary.  As a result, we then untension the fiber in 20 $\mu$m increments until the fiber buckles.   We observe the buckling process (the fiber bending inwards and then straightening under the inward force from the motors) with a 2 MP USB digital microscope mounted orthogonally to the flame above the center of the fiber, see 17 in Fig.~\ref{fig:pullingrig}.

The buckling process results in no loss in transmission and no thinning visible through the optical microscope upon introducing the fiber to the flame  within the microscope's resolution of $\pm2~\mu$m.   

\begin{figure}[H]
\begin{center}
\includegraphics[width=\linewidth]{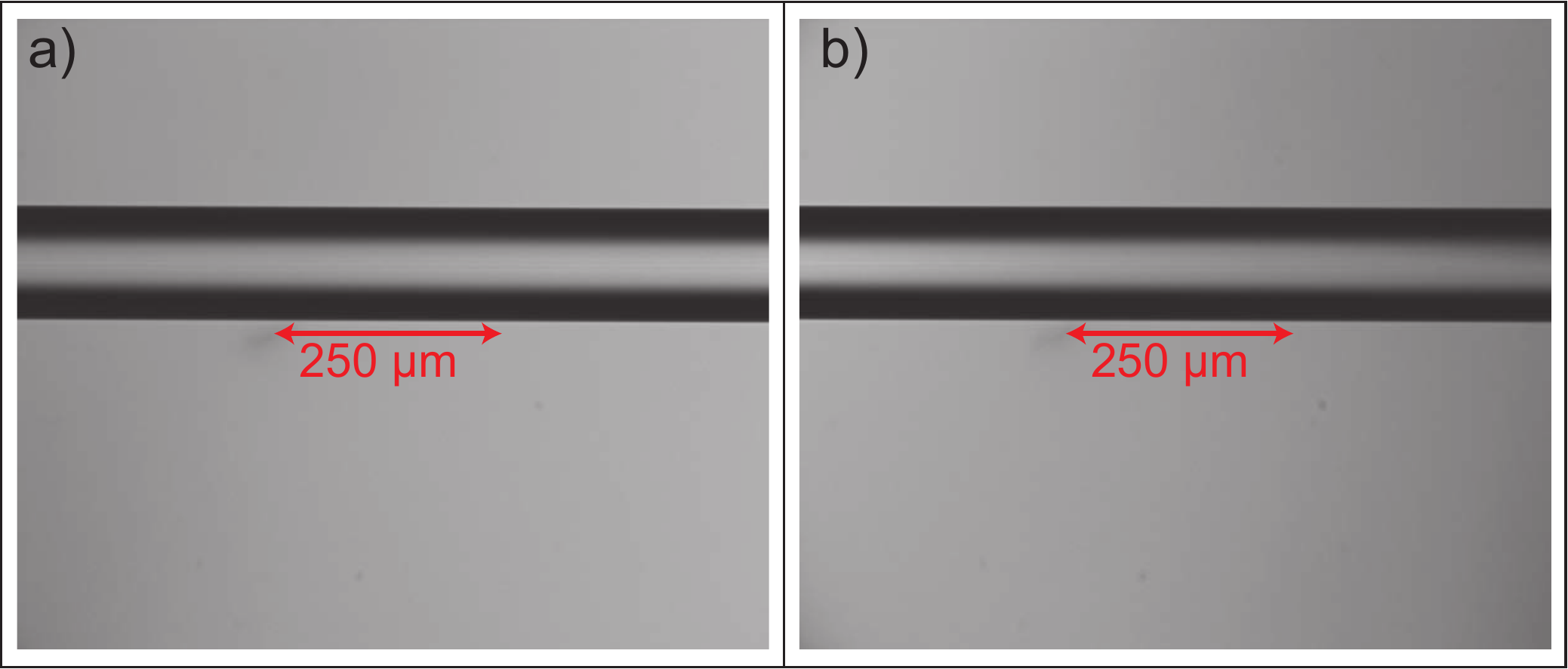}
\caption{Images of the section of fiber held directly next to the left (a) and right (b) fiber clamps respectively. The two images are separated by 3 cm. }
\label{fig:fiberalignment}
\end{center}
\end{figure}

Once the fiber is properly tensioned, we align the fiber such that the sections of fiber directly next to the left fiber clamp and right fiber clamp are equidistant from the optical microscope and at the same height.  We translate each section of fiber in front of the optical microscope using the fiber motors, see Fig.~\ref{fig:pullingrig}, and  align the height and focus of each fiber section using the micrometers attached to the flexure stage until both images overlap. If we see a sag in the fiber caused by the buckling we carefully retention the fiber in $5\ \mu$m steps until the fiber is straight as in Fig.~\ref{fig:fiberalignment}.  The microscope objective has a 3 $\mu$m depth of field, so by matching the diameter of the lensed light from the cladding and core  we ensure the v-grooves of the fiber clamps on each motor are equidistant from the camera, and therefore the nozzle of the flame. This  alignment is on the order of micrometers over a length of centimeters.  Once the images overlap, see Fig.~\ref{fig:fiberalignment}, the fiber is ready to be pulled.

\section{Results}\label{sec:res}

Here we discuss the results obtained from following the procedures outlined in the previous sections.  We present details on the transmissions achieved by following cleaning and alignment procedures.  Finally, we detail methods to aid in understanding the entire modal evolution during the fiber pull as a final check on the quality of the nanofibers we produce.

\subsection{Transmission}\label{sec:transmission}
\begin{figure}[H]
\begin{center}
\includegraphics[scale = 0.62]{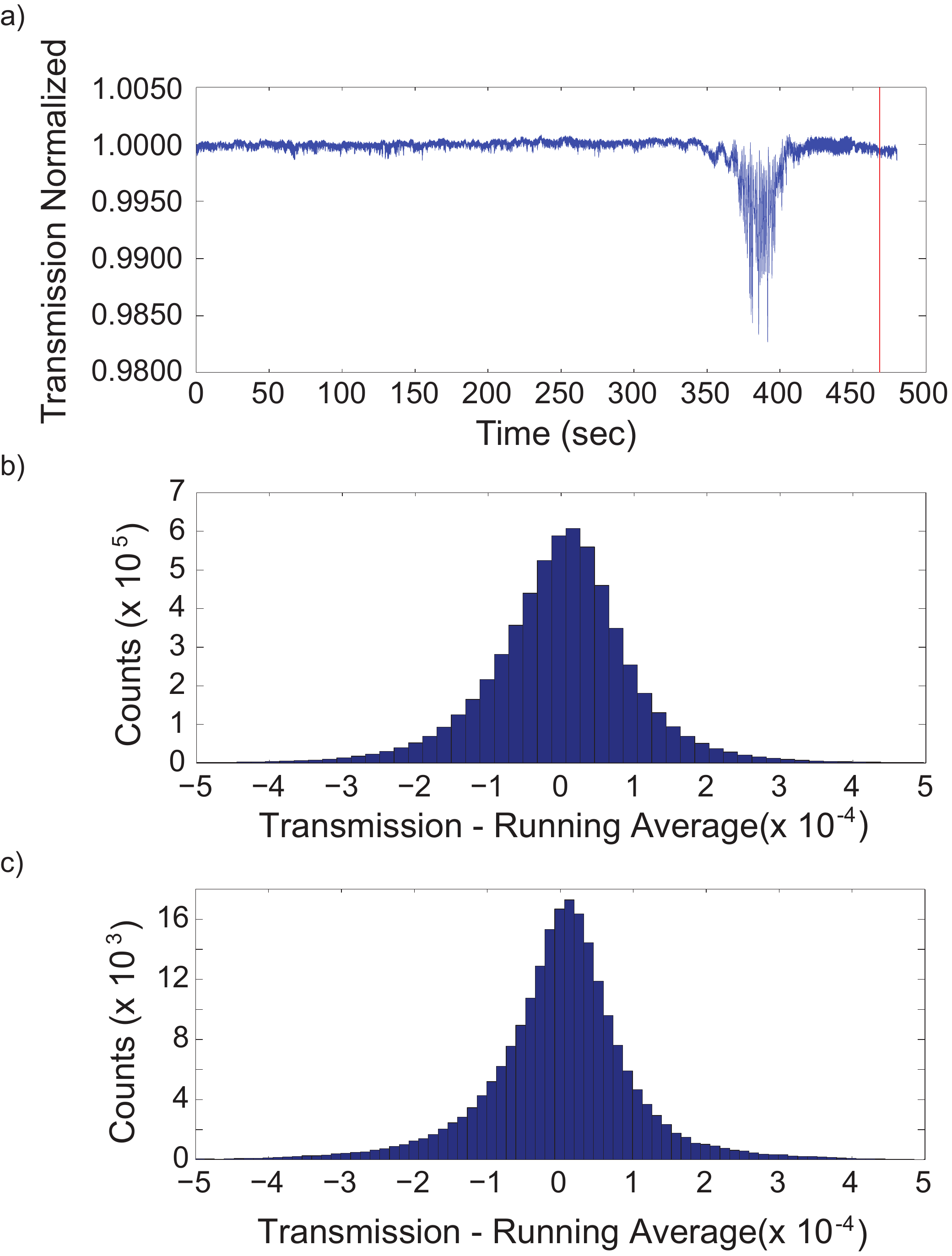}
\caption{a) The normalized transmission as a function of time during the pull through a nanofiber with an angle of 2 mrad to a radius of 6 $\mu$m and exponential profile to a final waist radius of 250 nm.  The length of the waist is 5 mm.  The fiber has a transmission of 99.95 $\pm0.02$ \% or equivalently a loss of 2.6 $\,\times \,$ 10$^{-5}$ dB/mm. b) Histogram of data taken from the beginning of the pull before higher order mode excitation. c) Histogram of transmission data taken after the pull ended.}
\label{fig:transmission}
\end{center}
\end{figure}

Figure~\ref{fig:transmission}(a) shows the transmission as a function of time during the pull for an optical nanofiber with a 2 mrad angle taper to a radius of 6 $\mu$m and exponential profile to reach a final waist radius of 250 nm, with a fiber waist length of 5 mm.  We achieve a transmission of 99.95~$\pm$~0.02~\%, or a loss of 2.6 $\times\,10^{-5}$ dB/mm.  The error listed in the transmission is the standard deviation.  We see from Fig.~\ref{fig:transmission}~(b) and (c) a histogram of the data ranges we use to take the mean for the average value at the beginning of the pull and the average value at the end of the pull.  We obtain a standard deviation of 1.0~$\times10^{-4}$ for the data in Fig.~\ref{fig:transmission}~b) and 1.0~$\times 10^{-4}$ for the data in c).  This leads to a standard deviation of the final transmission of  2.3~$\times 10^{-4}.$  There are possible systematic errors such as drifts in the amplifier gain, but we expect them to be of the same order or smaller than the quoted standard deviation. 
The final transmission is determined by  taking the mean of the data after the pull ends, delineated by the red line in Fig.~\ref{fig:transmission}, and dividing by the value of the normalized signal at the beginning of the pull, which we must determine.  We take a cumulative average of the  transmission from the beginning of the pull until just before any higher order modes are excited, see Sec.~\ref{sec:spec}.  Using this we see that the transmission  steadies at 99.95 \%.  We find this a fair method because there is no detectable loss over this range and no beating between modes \ref{sec:dustaccumulation} in the signal since we have yet to excite any higher order modes, and by checking the cumulative average we show that the transmission listed is steady and a lower bound.

\begin{figure}[H]
\begin{center}
\includegraphics[width=\linewidth]{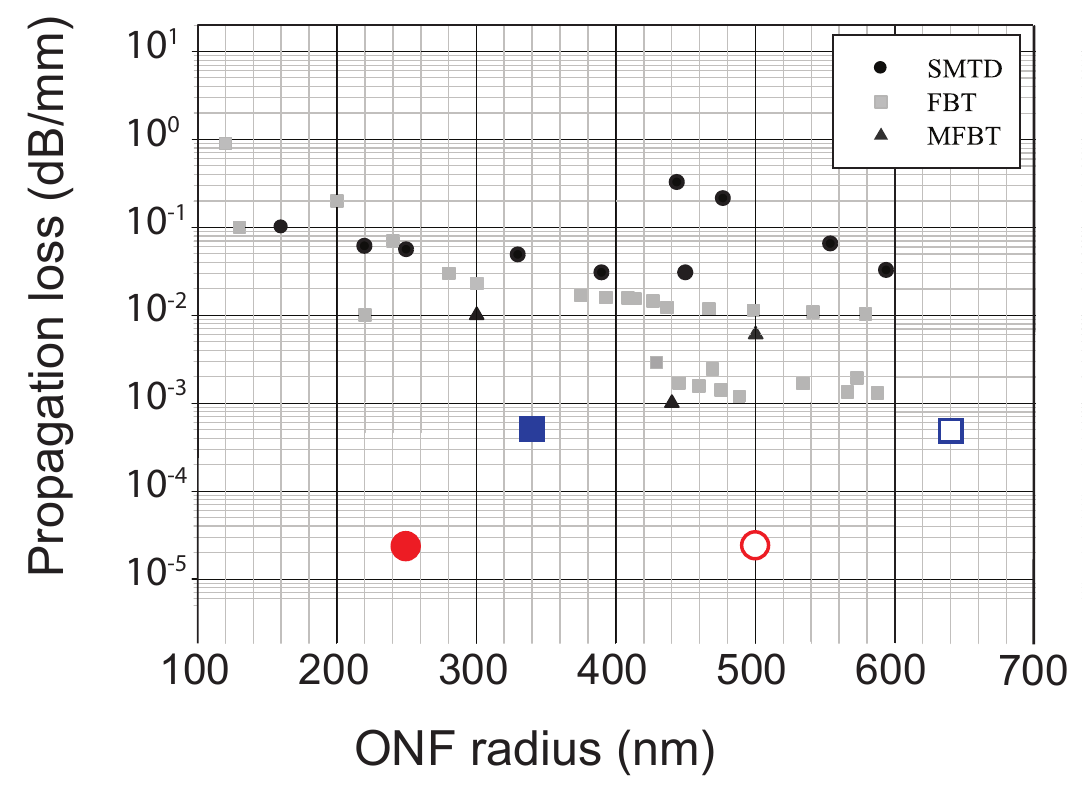}
\caption{Propagation loss as a function of optical nanofiber radius as compiled in Ref.\cite{Brambilla2010} edited to include our results for optical nanofiber loss in dB/mm.  The smaller, solid gray and black squares, circles, and triangles, represent previous results.  The solid red circle represents our result for loss launching the fundamental mode.  The open red circle scales the fundamental mode result to the effective radius to compare results with equivalent V numbers at 1550 nm.  The solid blue square represents our loss when launching the higher-order modes. Similarly the open blue square scales the result to the effective radius.}
\label{fig:brambrev}
\end{center}
\end{figure}

Previous work has focused on telecom light at 1550 nm.   In Fig.~\ref{fig:brambrev} we plot the propagation loss as a function of optical nanofiber radius for different pulling techniques as compiled in Ref.\cite{Brambilla2010} and references therein. We extend the axes  to overlay our results. Figure~\ref{fig:brambrev} shows that the lowest loss for previous work on the fundamental mode is on the order of 10$^{-3}$ dB/mm at 1550 nm, with final radii of between 440-600 nm.  Our result for the fundamental mode has a loss of 2.6 $\,\times \,$ 10$^{-5}$ dB/mm when the loss is taken over the entire 84 mm stretch. If the loss is only attributed to the 5 mm waist this becomes 4.34$\,\times \,$ 10$^{-4}$ dB/mm.    These results mark an improvement of two orders of magnitude over previous work \cite{Brambilla2010, Brambilla2004, Leon-Saval2004}.  For higher-order mode pulls using SM1500 fiber with an initial diameter of 50 $\mu$m,  the taper angle was 0.4 mrad until a radius of 6 $\mu$m and then had an exponential profile until reaching a uniform waist radius of 280 nm.  Here we achieved a loss of 5 $\times 10^{-4}$ dB/mm, when taken over the entire stretch, which represents less loss than the previous results for fundamental mode launches.   

Since the V number is proportional to the  fiber radius divided by the input wavelength we find it fair to compare our results at a wavelength of  780 nm  to the results in Fig.~(\ref{fig:brambrev})at 1550 nm by scaling our final radius  by a factor of 2.  The solid red circle and blue square in Fig.~\ref{fig:brambrev} represent the actual radius of the pull while the open red circle and blue square are designed to scale our results to equivalent V numbers for inputs at 1550 nm and represent an effective radius.  This means our effective final radius for the fundamental mode is ~ 500 nm and for the higher modes  ~ 640 nm.  This ultra-high transmission is reproducible to better than ~ 1\% over time with the same fiber, when following the cleaning and alignment procedure outlined in Sec.~\ref{sec:pp}.  
 
Using a numerical Maxwell's equations solver, FIMMPROP\cite{FIMMPROP}, we simulate the expected transmission through a fiber with the same profile as in our pulls.  We find the expected transmission to be 99.97\%\cite{Ravets2013},  through a $one-sided$ taper profile matching the 2 mrad pull depicted in Fig.~\ref{fig:transmission}, which is consistent with our experimental result that measures the transmission through the entire nanofiber.  Furthermore, when launching the next family of modes  through the fiber the FIMMPROP simulations were well-matched to the achieved transmissions\cite{Ravets2013a}. This implies that we are not limited by the pulling apparatus.

\subsection{Spectrogram Analysis}\label{sec:spec}

\begin{figure}[H]
\centering
\includegraphics[width=\linewidth]{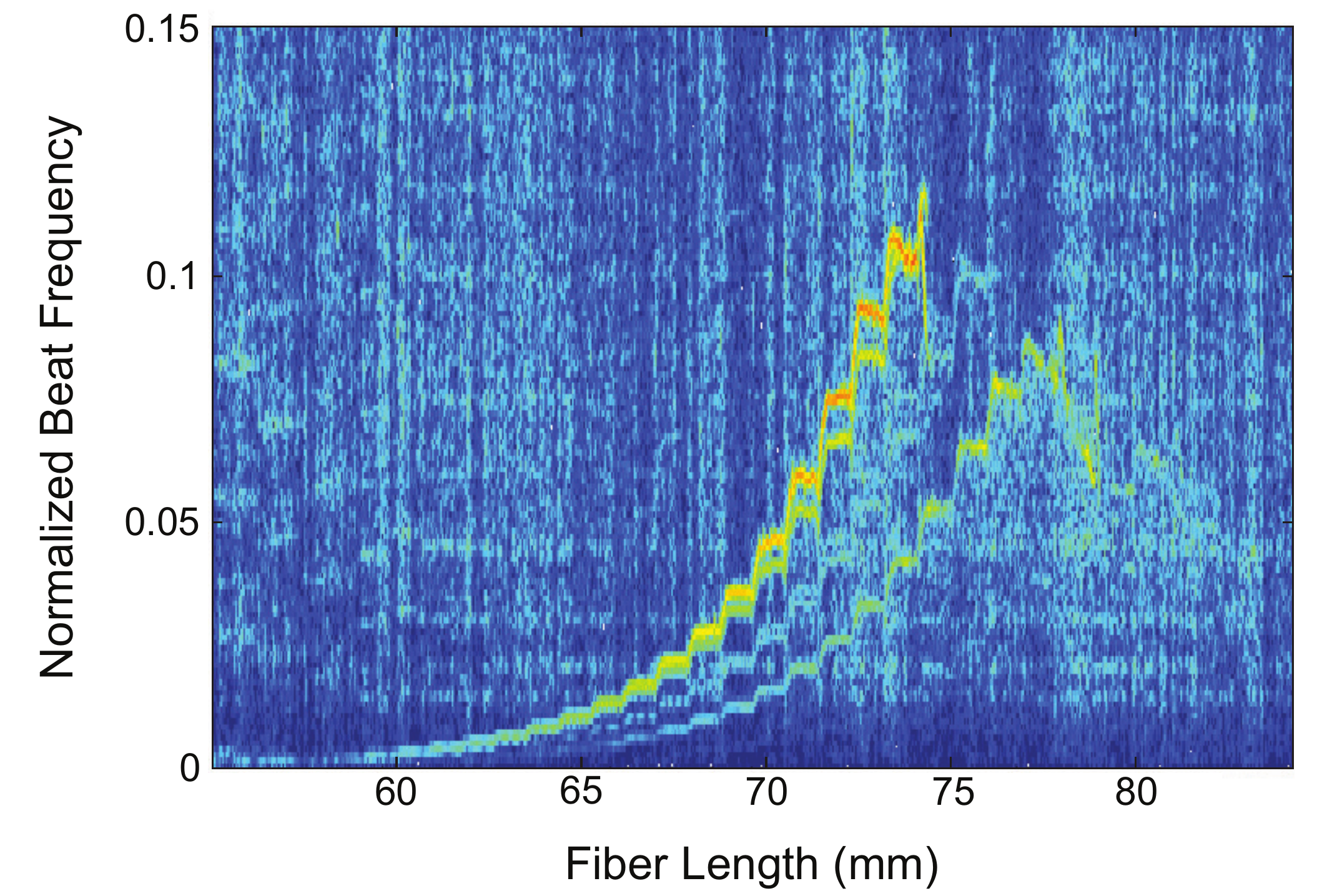}
\caption{Spectrogram of the transmission data from Fig.~\ref{fig:transmission}.  The excited curves correspond to higher-order modes of the same symmetry as the fundamental mode: $EH_{11}$, $HE_{12}$, and $HE_{13}$.}
\label{fig:spec}
\end{figure}

We analyze the quality of the nanofiber using a spectrogram, a short-time Fourier transform of the transmission data, also sometimes referred to as the Gabor Transform.  The spectrogram allows us to extract the entire modal evolution in the nanofiber during the pull.    Each curve corresponds to the evolving spatial beat frequency between the fundamental mode and excited modes propagating in the fiber, while the contrast corresponds to the energy transferred from the fundamental mode.    Following this we can use theory and simulation to identify all modes that are excited during the pull.  A detailed description, with full theoretical background, can be found in Ref.~\cite{Ravets2013}.

Figure~\ref{fig:spec} is a spectrogram of the transmission data from Fig.~\ref{fig:transmission}.  We see that for a successful 2 mrad pull with SM800 fiber we expect to observe a few higher order mode excitations. If modes are excited that are asymmetric to the fundamental mode, we know the cylindrical symmetry of the fiber was broken during the pulling process\cite{Ravets2013}, which can aid in identifying and fixing the error in the pulling apparatus.

It is worth noting that the modal excitation remains in the family of the same symmetry as the fundamental mode.  To the best of our knowledge, this is the first report of modal excitation that remained purely in the symmetric family of modes.  Previous work has seen asymmetric excitations to the $TE_{01}$, $TM_{01}$, and $HE_{21}$ modes\cite{Ding2010, Orucevic2007}.  In Sec.~\ref{sec:dustaccumulation}, we demonstrate this is not the case for an uncleaned fiber.

\section{Power Measurements}\label{sec:PM}

Once the pull is complete the fiber is transferred to a HV chamber beneath a HEPA filter.  Without keeping the fiber in a clean environment the transmission will degrade \cite{} as dust accumulates on the surface of the nanofiber waist and taper, which will cause the fiber to break under high powers in vacuum due to heating \cite{Wuttke2013}.   


To prevent the leakage current from impinging on the nanofiber, we avoided direct line of sight between the ion pump and the nanofiber by placing the ion pump on an elbow.  Then between the elbow and the ion pump we placed a grounding mesh to prevent the electric field from penetrating past the mesh.   With this arrangement a 250 nm radius nanofiber has withstood the application of more than 400 $\pm$ 12 mW from a Ti:Sapphire laser at 760 nm in HV conditions.

\section{Conclusion}\label{sec:con}

We provide the necessary procedures to clean, prepare, and pull an ultrahigh transmission nanofiber in a reproducible way.  The work is validated through microscopy, and we present the transmission results of a standard 2 mrad pull yielding a transmission of 99.95 $\pm$ 0.02\% or loss of 2.6~$\times \, 10^{-6}$~dB/mm, an improvement of two orders of magnitude for the fundamental mode.  When launching higher-order modes we have losses of 5 $\times 10^{-4}$ dB/mm.  The transmission results are in excellent agreement with transmission simulations, implying the limiting factor in transmission comes from a lack of pre-pull cleanliness.  We provide a detailed cleaning protocol, which greatly improves the reproducibility for ultrahigh transmission fibers and produces the first recorded tapers without asymmetric modal excitation. We provide evidence that the pre-pull cleanliness is critical to achieving ultrahigh transmission nanofibers.  These fibers can achieve efficient guidance with short, controllable taper lengths and are usable for various atomic physics applications.  During the manuscript writing process we became aware of similar independent work\cite{Ward2014}.


%
%

%

\begin{acknowledgments}
Work supported by National Science Foundation of the USA through the Physics Frontier Center at the Joint Quantum Institute, Army Research Office Atomtronics MURI, and S. R. thanks the Fulbright Foundation for support.  We acknowledge the support of the Maryland NanoCenter and its NispLab. We would like to thank Fredrik Fatemi and Guy Beadie for their major contributions to the higher-order mode studies and  Prof. A. Rauschenbeutel for his support and interest in this project.  \end{acknowledgments}

\section*{Appendix A}\label{sec:alg} 
We pull our fibers using a flame brushing technique \cite{Bilodeau88, Birks1992, Warken2007,Warken2007b}. A section of fiber, less than a millimeter in length, is brought to its softening point using a clean oxyhydrogen flame and then pulled by two high-precision motors.\\

Our algorithm\footnote{The program is available at the Digital Repository of the University of Maryland DRUM}, based on the work of the originally Mainz and currently Vienna group\cite{Warken2007b},  calculates the trajectories of the motors needed to produce a fiber with the desired final radius, length of uniform waist, and taper geometry. The tapers are formed by a series of small sections that are well approximated by lines, allowing us to form a linear taper with a given angle down to a radius of 6 $\mu$m, which connects to an exponential that smoothly reduces to a submicron radius, typically 250 nm. The slope of the linear section generally varies between 0.3 and 5 mrad. Our algorithm divides the pull into steps defined by their pulling velocity and the traveling length of the flame. We recursively calculate the parameters, starting from the desired final radius, $r_w$,  until reaching the initial radius, $r_0$.  The full details and code can be found at Ref.~\cite{Note1}.  

\subsection{Motor Control} 
The model produces a velocity profile that is a square wave in time.  Experimentally, we approximate the square wave in three parts:
\begin{enumerate}
\item A ramp up to $v_{b,n} \pm v_{f,n}/2$
\item A constant pull velocity equal to $v_{b,n} \pm v_{f,n}/2$
\item A ramp down in velocity to zero.
\end{enumerate}

Where  $v_{b,n}$ is the velocity of the flame in step n and $v_{f,n}$ is the velocity that the fiber motors move apart.  The  addition of $v_{b,n}$ arises from the transformation to the rest frame of the flame. 
Typically, $v_{b,n}$  is an order of magnitude greater than $v_{f,n}$.  When transforming to the rest frame of the flame, both motors move in the direction the flame would have swept in that step. The motor whose pull velocity is in the same direction as the flame motion will lead while the other motor will lag. We have verified this sequence using the encoder of the motor that allows us to record the trajectory of the motors and by looking at the output of a Michelson interferometer with one arm spanning the two motorized stages.

\subsection{Measurement of the flame width}\label{sec:flamemeasurement}

One experimental parameter fundamental to the algorithm is the effective size of the flame, $L_0$, which corresponds to the zone of the fiber inside the flame that melts and thins during the pulling process. The softening point for the fused silica used by Fibercore for the SM800 fiber occurs at 1585$^{\circ}$ C. The best way to evaluate this is to measure the impact of the flame on the fiber, since our flame cannot be observed by eye.

Working with reproducible conditions requires that we fix the working distance between the fiber and the nozzle. As a consequence, the fiber is always at the same spot inside the flame and always sees the same distribution of temperature. We check the distance with a microscope before each pull and fix it to 400 $\pm$ 50 $\mu$m.

We measure $L_0$ by fixing the flame and letting both motors move apart at a constant velocity. Conservation of volume leads to an exponential profile with a waist of length $L_0$, and the radius profile is given by :
\begin{align}\label{eqn:uni}
r_w = r_0 \exp{\left( -\frac{t_h v_f}{2 L_0} \right)} ,
\end{align}
where $t_h$ is the heating time and $r_0$ the unmodified radius of the fiber.  

We use our imaging system (15 in Fig.~\ref{fig:pullingrig})to measure the radius of the waist of the fiber for different values of $v_f t_h$, and fit $\ln{(r_0/r_w)}$ to extract $L_0$.

The measurement consists of fixing the pulling velocity at 0.05 mm/s,  varying the heating time from 2 to 32 s, and then measuring the final radius of the waist. We limit ourselves to times less than 40 seconds to stay within the  2 $\mu$m resolution of our imaging system.

We characterize the size of the flame by plotting $\ln{(r_0/r_w)}$ as a function of $v_f t_h$ we obtain a fit 
with a reduced $\chi^2$ of 1.07 that yields $L_0 = 0.753 \pm  0.014$ mm.  This parameter should be checked from time to time as the pulling apparatus is used since it can vary by a small amount.


Measuring the length of the waist or fitting the profiles of the taper to an exponential are less accurate methods than the above procedure because Eq.~\ref{eqn:uni} assumes a uniform hot zone, $L_0$.  In this measurement we keep the flame fixed, which means that our hot zone is not uniform.  During the actual pulling procedure we sweep, which creates an effective uniform hot zone.  Here, the section of fiber located at the central point of the flame is thinned the most, as a result it is more accurate to  measure the profile of the fiber after tapering and find the smallest radius to extract the value of $L_0$.

\subsection{Microscopy validation}
\begin{figure}[H]
\begin{center}
\includegraphics[width=\linewidth]{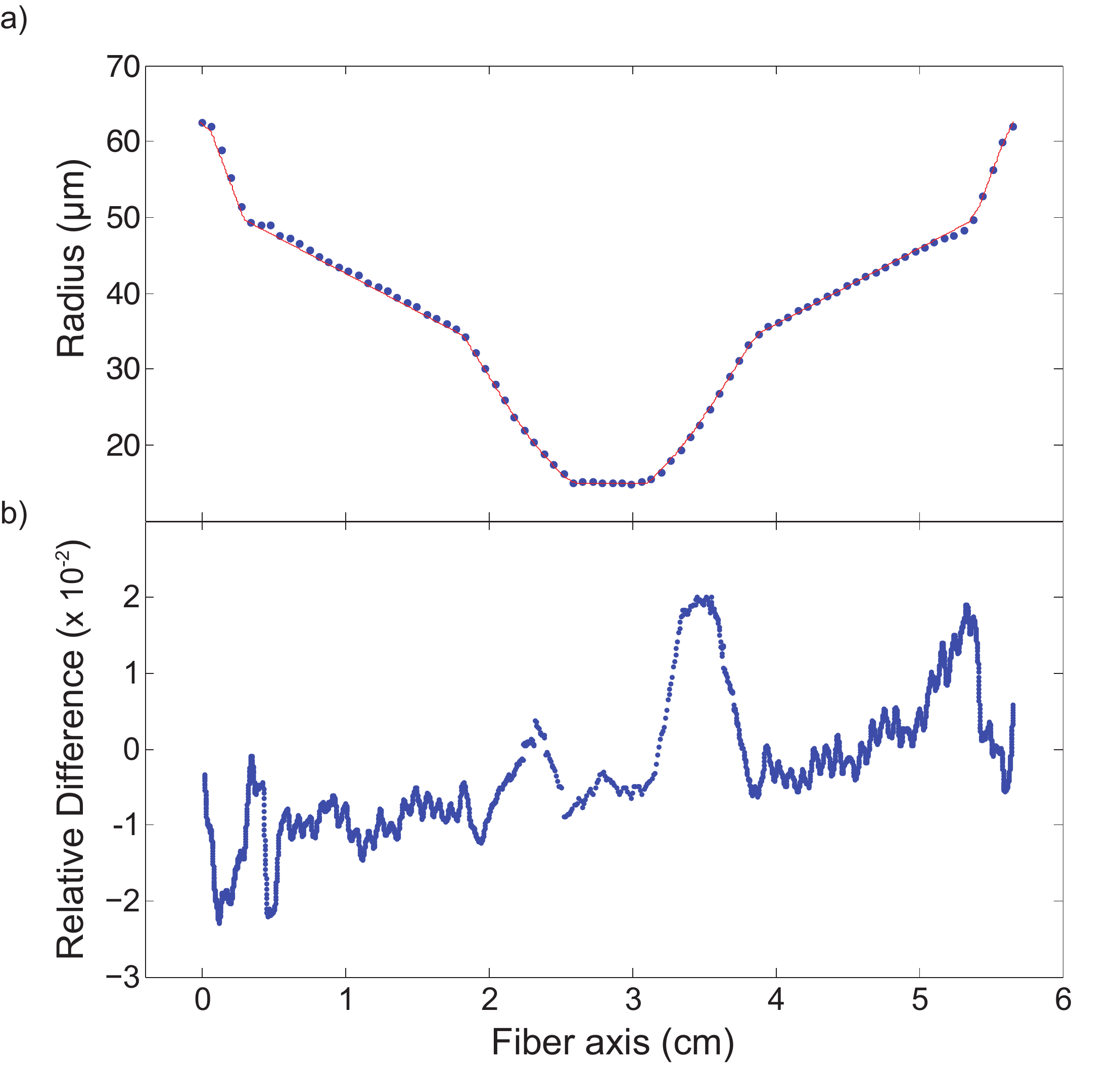}
\caption{Profile of a multiple-angled, linearly tapered fiber.  (a)  The blue dots are measurements taken using the optical microscope and the red line represents the profile shape from the fiber tapering simulation. The pull was for a final radius of 15 $\mu$m. The taper profile was designed to have three angles, 5, 2, 3 mrad,  that taper down to radii of 50, 35 and 25 $\mu$m, respectively.  The error in each measurement is dominated by a systematic error of~$\pm$2.5 $\mu$m. (b) The relative difference between the expected profile and the measured profile with an  RMS value of 0.0187.}
\label{fig:profile}
\end{center}
\end{figure}

We validate the accuracy of our simulation of the expected fiber profile using both an $in\ situ$ optical microscope and a scanning electron microscope (SEM).
Figure~\ref{fig:profile}~(a) shows the measured (blue markers) and simulated profiles (red lines) of a fiber taper imaged optically.  The taper profile is designed to have three angles, 5, 2, and 3 mrad,  that taper down to radii of 50, 35 and 25 $\mu$m, respectively. An exponential profile smoothly links the radius of 25 $\mu$m down to the the final radius of 15 $\mu$m.  The final radius is chosen to be well above the resolution of our optical microscope.   The length of the uniform waist is chosen to be 5 mm long.  Figure~\ref{fig:profile}(a) is a compilation of optical microscope  images taken of the entirety of the tapered fiber. An edge finding technique then measures the profile of the fiber at different cuts.  The error in the measured radius is dominated by a systematic error of  ~$\pm$2.5 $\mu$m due to the finite resolution of the imaging system.  We first  use an image of the unmodified fiber, which has a diameter of 125.1 $\mu$m, to determine the pixel to micron conversion.  The number of pixels measured for an unmodified fiber has an error of a few pixels as a result of the resolution of the optical microscope.   We then binarize the gray levels of the pixels and choose a threshold such that the diameter of the unmodified fiber matches the pixel count from the previous measurement.  The edge finding technique itself has an error of about 0.5 pixels for a flat length of fiber resulting from the binarization process.
Figure~\ref{fig:profile}~(b) displays the relative difference between the measured image radius and the simulated radius normalized to the expected radius.  The largest deviation is slightly larger than 2\%, while the RMS value is 0.0187.  This verifies the accuracy of our algorithm and pulling apparatus for larger radius tapers.   

\begin{figure}[H]
\begin{center}
\includegraphics[width=\linewidth]{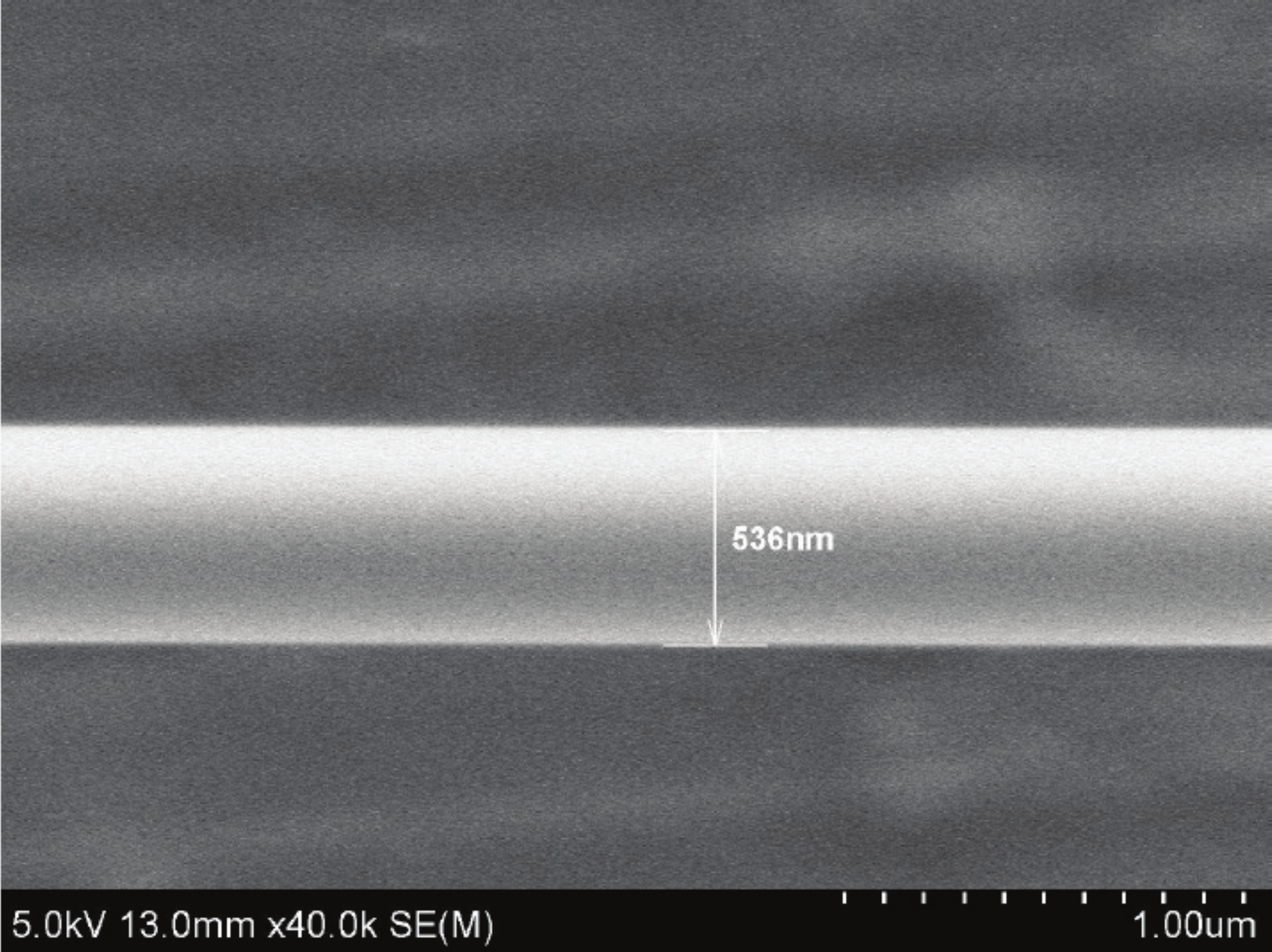}
\caption{A SEM image taken in the NISP lab at UMD.  We measure a radius of 536  $\pm$ 12 nm,  the expected diameter of the waist is 500 nm.}
\label{fig:SEM}
\end{center}
\end{figure}

We  use a SEM to measure the nanofiber profile below a micrometer to verify that our nanofibers truly achieve the desired diameter.  Figure~\ref{fig:SEM}, shows a SEM image of a nanofiber, coated with graphite, with an expected diameter of 500 nm, and a measured diameter of 536 $\pm$ 12 nm.  The error is systematic, coming from the scaling factor associated with the SEM calibration.   We attribute this small disagreement to thermal forces that push the fiber away from the nozzle at the end of the pull when the fiber is thin.  We could compensate for this in the algorithm by adjusting the effective hot zone as the fiber tapers, but we have not found it necessary to do so.


\section*{Appendix B} \label{sec:dustaccumulation}          

Any particulate accumulation on the optical fiber before the pull begins will compromise the quality of the optical nanofiber:  it will degrade the transmission, excite higher order modes, change the modal evolution, and scatter light.  If any particulate accumulates on the fiber before the pull, the maximum possible transmission for a given taper geometry will not be achieved.  Using FIMMPROP, as described in \ref{sec:transmission}, we have a sense of what this ideal transmission is for a given geometry, and if our transmission deviates, it can generally be attributed to a lack of proper cleaning.  If the nanofiber environment is not clean or has a high humidity the transmission will decrease after a pull is finished\cite{Fujiwara2011}.  Furthermore, if any dust accumulates on the nanofiber surface, it will not withstand high powers under vacuum. 

\begin{figure}[H]
\centering
\includegraphics[width=\linewidth]{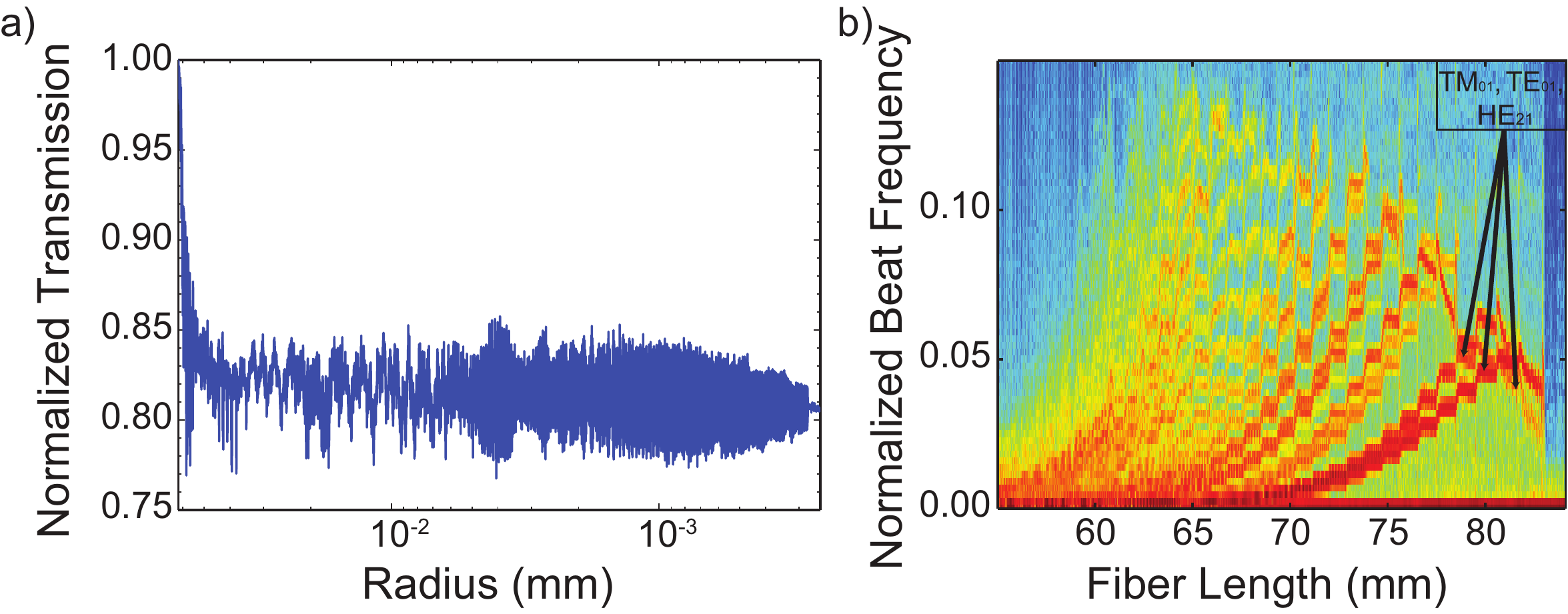}
\caption{(a) Transmission signal for an uncleaned 2 mrad tapered fiber. The  transmission of this fiber is  80.50 \%. (b) Spectrogram of the transmission data. We can distinguish the excitation of many higher order modes.  Of special interest are the curves identified, the asymmetric $TE_{01}$, $TM_{01}$, and $HE_{21}$ modes. 
}\label{dirt}
\end{figure}

If the fiber is not properly cleaned before pulling, the final transmission can vary by a few percent.   Figure~\ref{dirt} displays the extreme case of mechanically stripping the buffer and not cleaning the fiber at all before pulling.  Here, the transmission is only 80.5\% for a 2 mrad taper down to $r_w=250$ nm, leading to more than a 19\% loss in transmission when compared to a properly cleaned fiber.  The spectrogram in Fig.~\ref{dirt} (b) shows excitation to excited asymmetric mode: $TE_{01}$, $TM_{01}$, and $HE_{21}$, identified by arrows, that were not present when the fiber was properly cleaned.  It is further interesting that there is more energy transferred to these asymmetric modes than any other modes. 

Before every pull, we follow the cleaning procedure described in Sec.~\ref{sec:cleaningprocedure}.  After imaging the fiber, we decide whether or not we should start the pull or restart the cleaning process.   We restart if anything is  obstructing the light traveling through the fiber reaching the CCD.  When there is particulate attached on top of or below the fiber, we use a wipe with methanol and remove it.  If there is nothing observable within the resolution of the optical microscope we proceed with the pull.  When we do not follow these criteria the reproducibility in the transmission will change by a few percent.  When we apply this cleaning method, the variability between runs is better than 1~\%.

The origin of the particulate can come in various forms: remnants of plastic buffer, solvent evaporate, or any small particulate floating in the air.  We believe the most common source to be the buffer.  Since we use a mechanical fiber stripper to remove the buffer, micro or macroscopic pieces of buffer remain on the fiber after stripping.  We apply wipes to remove the buffer remnants. This removal process can be imperfect because mechanical strippers are not designed to make contact with the actual glass of the fiber.  Buffer remnants are a particularly insidious form of particulate.  The plastic is generally designed with a higher index of refraction than the cladding to help remove cladding modes.  If the buffer remains, we believe it may burn into the fiber.  This higher index irregularity can lead to an excitation of higher order modes.

During the pull there may be signatures that the fiber was not properly cleaned.  These take the form of large losses in transmission and the excitation of higher order modes.  If there is initially loss or beating in the transmission this is a sign that the fiber was not properly cleaned; this is displayed in Fig.~\ref{dirt}(a).  The fiber starts single mode in the core and therefore there should be no beating between modes and negligible losses in the initial pulling process, before the fundamental mode becomes a cladding mode, in which the tapering process reduces the effective index of refraction to and then below the index of refraction of the cladding. 

\begin{figure}[H]
\centering
\includegraphics[width=\linewidth]{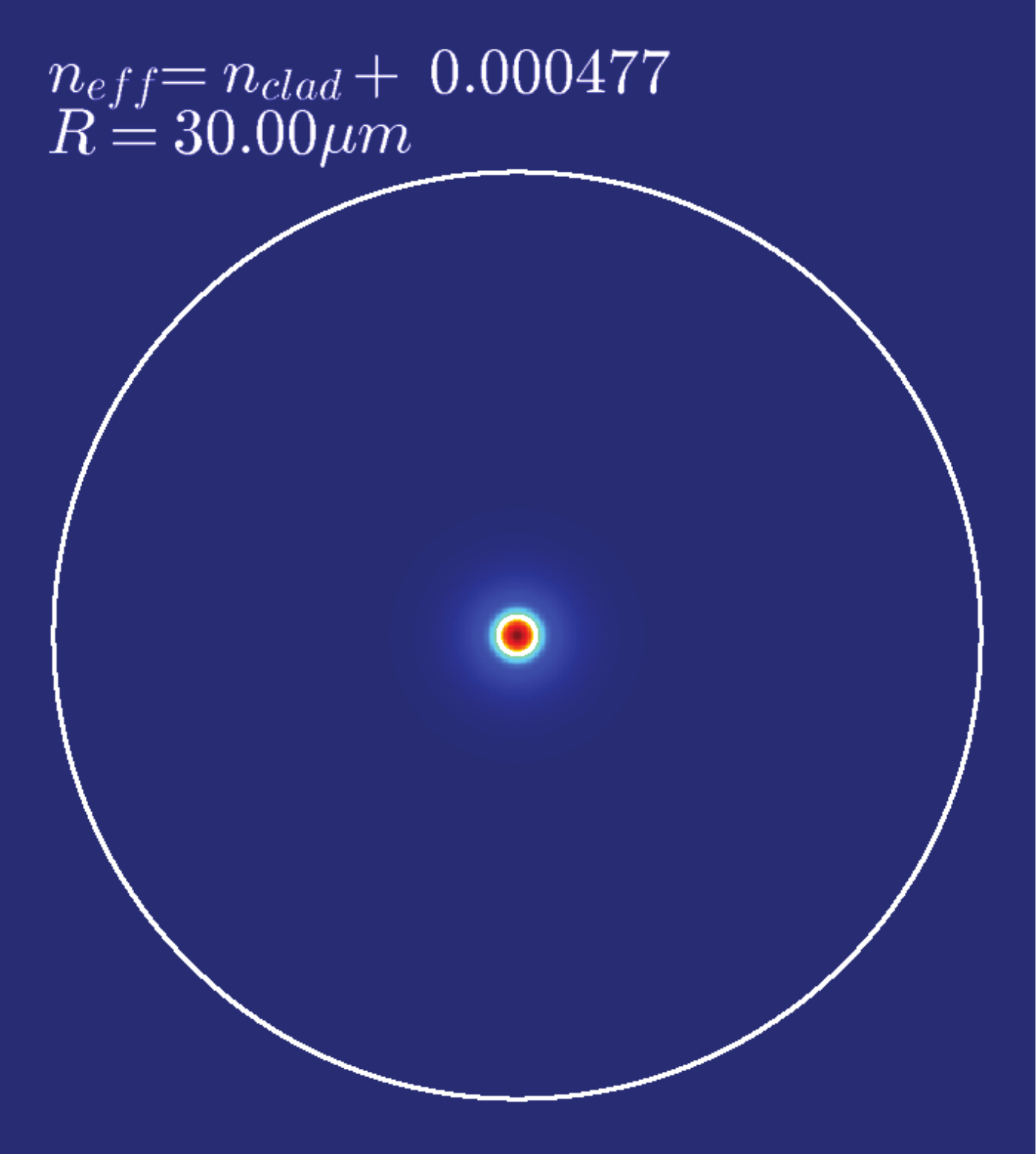}
\caption{The transition of the fundamental mode from a core to a cladding mode.  The intensity of each frame is renormalized for visualization.  The modes are obtained using the numerical solver FIMMPROP.
}\label{mov:transition}
\end{figure}

As the fiber continues to taper and the effective index of refraction of the fundamental mode approaches the index of refraction of the cladding, the mode begins to leak from the core.  This is when higher order mode excitation can occur.  For the SM800 fiber used in this study, the transition occurs at a radius of 19.4 $\mu$m.  This transition is captured by Media~ 1 in Fig.~\ref{mov:transition}.  Here, we see that as the radius reduces, the effective index of refraction approaches the index of refraction of the cladding and the mode begins to leak from the core.   If the beating between higher order modes occurs before this point, this is evidence that the fiber was damaged.  A nanofiber with a 2~mrad geometry, if handled properly, typically excites only three higher order modes: $EH_{11}$, $HE_{12}$, and $HE_{13}$ \cite{Ravets2013}.  In  Fig.~\ref{dirt}(b), each curve corresponds to beating between different modes and we can identify more than twenty excited modes as a result of the buffer remnants.  

Furthermore, any dust on the nanofiber will cause it to break under high power in vacuum. The cleanroom environment and the cleanliness of our pulling and transfer procedures allow us to achieve nanofibers that transmit more than 400 $\pm$  12 mW in an HV environment.

We believe that chemically removing the buffer could be beneficial to the fiber transmission.  Chemical removal can lead to less mechanical damage to the fiber and properly remove all of the buffer.  This is not a critical issue, since our transmission is in good agreement with simulations from FIMMPROP, but it could improve reproducibility and ease the cleaning process.

\bibliographystyle{aipauth4-1}


%
\end{document}